\newcommand{\be}{\begin{equation}}\newcommand{\ee}{\end{equation}}
\newcommand{\bea}{\begin{eqnarray}}\newcommand{\eea}{\end{eqnarray}}
\newcommand{\nn}{\nonumber}\newcommand{\p}[1]{(\ref{#1})}
\newcommand{\lb}[1]{\label{#1}}
\newcommand\s{\scriptscriptstyle}

\newcommand\q{\quad}
\newcommand\qq{\quad\quad}
\renewcommand\={\ =\ }

\newcommand\cW{{\cal W}}
\newcommand\bcW{\bar{\cal W}}

\newcommand\stc{\stackrel{\star}{,}}

\newcommand\olp{\overleftarrow{\partial}}
\newcommand\orp{\overrightarrow{\partial}}
\newcommand\olD{\overleftarrow{D}}
\newcommand\orD{\overrightarrow{D}}
\newcommand\olbD{\overleftarrow{\bar D}}
\newcommand\orbD{\overrightarrow{\bar D}}

\newcommand\tpa{\theta^{+\alpha}}

\newcommand\tma{\theta^{-\alpha}}

\newcommand\tka{\theta^{\alpha}_k}

\newcommand\tib{\theta^{\beta}_i}

\newcommand\btka{\bar\theta^{\da k}}
\newcommand\btkb{\bar\theta^{\db k}}

\newcommand\btja{\bar\theta^{\da j}}

\newcommand\btpa{\bar{\theta}^{+\dot{\alpha}}}

\newcommand\tp{\theta^+}

\newcommand\btp{\bar\theta^+}

\newcommand\ada{{\alpha\dot{\alpha}}}
\newcommand\adb{{\alpha\dot{\beta}}}

\newcommand\ab{{\alpha\beta}}
\newcommand\ba{{\beta\alpha}}

\newcommand\da{{\dot{\alpha}}}
\newcommand\db{{\dot{\beta}}}

\newcommand\A{{\s A}}

\newcommand\R{{\s R}}

\newcommand\sL{{\s L}}

\newcommand\W{{\s W}}
\newcommand\Z{{\s Z}}

\newcommand{\pp}{{\s ++}}
\newcommand{\m}{{\s --}}

\newcommand{\Dp}{D^{\pp}}
\newcommand{\Dm}{D^{\m}}

\newcommand{\Vp}{V^\pp}
\newcommand{\Vm}{V^\m}

\newcommand{\Dpa}{D^+_\alpha}

\newcommand{\Dpb}{D^+_\beta}

\newcommand{\bDpa}{\bar{D}^+_{\dot{\alpha}}}
\newcommand{\bDpb}{\bar{D}^+_{\dot{\beta}}}
\newcommand{\bDma}{\bar{D}^-_{\dot{\alpha}}}

\def\sfrac#1#2{{\textstyle\frac{#1}{#2}}}
\def\ha{\frac12}
\def\sha{\sfrac12}

\def\e{\mbox{e}}
\def\ii{\mbox{i}}
\def\diff{\mbox{d}}

\documentclass[12pt]{article}
\usepackage{amsmath,amssymb}

\begin{document}
\begin{flushright}
hep-th/0402062 \\[5mm]
ITP--UH--04/04 \\
[5mm]
February, 2004\\
[1.5cm]
\end{flushright}
\def\thefootnote{\fnsymbol{footnote}}
\begin{center}
{\Large\bf
Non-anticommutative N=(1,1) Euclidean Superspace\footnote{To be published
in Proceedings of the International workshop "Supersymmetries and quantum
symmetries", July 24-29, Dubna, 2003}}
\vspace{1.5cm}

{\large\bf
Evgeny Ivanov$\;{}^a$, \ Olaf Lechtenfeld$\;{}^b$, \ Boris Zupnik$\;{}^a$ }
\\
\vspace{1cm}

${}^a$ {\it Bogoliubov  Laboratory of Theoretical Physics, JINR,
141980 Dubna, Russia;} \\
{\tt eivanov, zupnik@thsun1.jinr.ru}

\vspace{0.3cm}
${}^b$ {\it Institut f\"ur Theoretische Physik, Universit\"at Hannover,
30167 Hannover, Germany; }\\
{\tt lechtenf@itp.uni-hannover.de}
\end{center}

\begin{abstract}
\noindent
We study deformations of four-dimensional $N{=}(1,1)$ Euclidean 
superspace induced by non-anticommuting fermionic coordinates.
We essentially use the harmonic superspace approach and consider 
nilpotent bi-differential Poisson operators only, which
generalizes the recently studied chiral deformation of $N{=}(\ha,\ha)$ 
superspace. We present non-anticommutative Euclidean analogs of $N{=}2$ 
Maxwell and hypermultiplet off-shell actions. The talk is based on the 
paper {\tt hep-th/0308012}.

\end{abstract}
\setcounter{equation}0

\noindent{\bf 1. Introduction}. This talk reports the results
of our recent paper \cite{ILZ} where we discuss
nilpotent deformations of $N{=}(1,1)$ Euclidean superspace.

Deformations of superfield theories are currently a subject of
intense study (see, e.g.~\cite{BrSc}--\cite{FLM}).
Analogously to noncommutative field theories on bosonic spacetime,
noncommutative superfield theories can be formulated in ordinary 
superspace by multiplying functions given on it via a star product 
which is generated
by some bi-differential operator or Poisson structure~$P$.
The latter tells us directly which symmetries of the undeformed (local)
field theory are explicitly broken in the deformed (nonlocal) case.

Generic Moyal-type deformations of a superspace are characterized by
a constant graded-antisymmetric non\-(anti)\-commutativity matrix
$(C^{AB})\,$. A minimal deformation of Euclidean $N{=}1$ superspace -- 
more suitably denoted as $N{=}(\ha,\ha)$ superspace -- was considered 
in a recent paper~\cite{Se}. For the chiral $N{=}1$ coordinates 
$(x^m_\sL,\theta^\alpha,\bar\theta^\da)$ the noncommutativity was 
restricted to
\be
\theta^\alpha\star\theta^\beta\=\theta^\alpha\,\theta^\beta+\sha C^\ab\ ,
\quad
\theta^\alpha\star x^m_\sL \= \theta^\alpha\,x^m_\sL\ ,\quad
x^m_\sL\star x^n_\sL \= x^m_\sL\,x^n_\sL\ ,
\lb{n2}
\ee
with $(C^\ab)$ being some constant symmetric matrix.
Note that the bosonic and the antichiral coordinates have undeformed
commutation relations with everyone, so $(C^{AB})$ is rather degenerate 
here. For functions $A$ and $B$ of $(x^m_\sL,\theta^\alpha,
\bar\theta^\da)$ the star product~\p{n2} is generated as
\be
A \star B \= A\,\e^P B \= A\,B + A\,P\,B + \sha A\,P^2 B
\ee
where the bi-differential operator is defined as
\be
P\=-\sha\olp_{\!\alpha} C^\ab \orp_{\!\beta}
\lb{P1}
\ee
and is nilpotent, $P^3\=0$. This provides a particular example of a 
deformed superspace. It retains $N{=}(\ha,0)$ of the original
$N{=}(\ha,\ha)$ supersymmetry because $Q_\alpha$ commutes with~$P$ while
$\bar{Q}_\da$ does not. It is natural to refer to the deformations 
generated by a nilpotent Poisson structure like~\p{P1} as {\it nilpotent 
deformations}.

While the preservation of chirality is the fundamental underlying 
principle of $N{=}1$ superfield theories~\cite{F0}, it is the power of 
{\it Grassmann harmonic analyticity\/} which replaces the use of 
chirality in $N{=}2$ supersymmetric theories in four dimensions~
\cite{GIK1,GIOS}. Therefore, it is natural to look for nilpotent 
deformations of $N{=}(1,1)$ Euclidean superspace which preserve this 
harmonic analyticity (perhaps in parallel with chirality). The basic aim 
of the present contribution is to describe, following ref. \cite{ILZ},  
such deformations and to give deformed superfield actions
for a few textbook examples of $N{=}2$ theories. We also discuss the role
of the standard conjugation or an alternative pseudoconjugation in 
Euclidean $N{=}(1,1)$ supersymmetric theories and their deformations.

Our main novel developments are the analysis of $N{=}(1,1)$
supersymmetry-breaking deformations in harmonic superspace and the 
construction of the relevant superfield models. Note that 
the supersymmetry-preserving deformations of $N{=}(1,1)$ superspace were 
considered in \cite{FL,KPT,BS} (without giving specific dynamical models) 
and, with making use of the harmonic superspace approach, also in ref. 
\cite{FS}.\footnote{This paper appeared in hep-th almost simultaneously 
with \cite{ILZ}.} The deformed $N{=}2$ harmonic superspace and field 
theory models in it are also addressed in a recent paper \cite{AIO}.

\vspace{0.3cm}
\noindent{\bf 2. Deformations of N=(1,1) superspace in a chiral basis}.
Our main goal is to generalize the nilpotent deformation of $N{=}1, D{=}4$ 
Euclidean superspace proposed in~\cite{Se}. In this deformation, one 
introduces non\-(anti)\-commutativity only for one half of the spinor 
coordinates. By construction, this deformation preserves the
chiral representations of $N{=}1$ supersymmetry.
The bi-differential operator of~\cite{Se} has the form~\p{P1} and acts 
on standard superfields $V(x^m_\sL,\theta^\alpha,\bar\theta^\da)$.

A a prerequisite, it is appropriate here to note that the 
(pseudo)conjugation properties of spinors in 4D Euclidean space with the 
group $Spin(4)$=SU$(2)_\sL\times\,$SU$(2)_\R$ are radically different 
from those in Minkowski space since left- and right-handed SU(2) spinors 
are independent. For $N{=}1$  Euclidean superspace to have the real 
dimension $(4|4)$ like its Minkowski counterpart, one is led to apply the 
following pseudoconjugation of the  SU$(2)_\sL\times\,$SU$(2)_\R$
spinor Grassmann coordinates (see e.g. \cite{Ma}):~\footnote{
We use the conventions $\varepsilon_{12}=-\varepsilon^{12}=
\varepsilon_{\dot{1}\dot{2}}=-\varepsilon^{\dot{1}\dot{2}}=1$,
$\sigma_m=(i\overrightarrow\sigma,I)$ for the basic quantities in the 
Euclidean space.}
\be
(\theta^\alpha)^*\=\varepsilon_\ab\theta^\beta\ ,\qq
(\bar\theta^\da)^*=\varepsilon_{\da\db}\bar\theta^\db\,,\qq (AB)^*=B^*A^*
\label{Oo}
\ee
where $A$ and $B$ are arbitrary superfields.
Here, the map  ${}^*$ is a pseudoconjugation which squares to $-1$ on any
odd $\theta$ monomial (and on the fermionic component fields) and to $+1$
on any even monomial (and on the bosonic component fields). So, when 
acting on bosonic fields, it can be identified with the standard complex
conjugation. It is straightforward to check that \p{Oo} is consistent 
with the action of the group $Spin(4)$ and preserves its irreducible 
representations. As an important consequence of the pseudoreality of 
spinor  coordinates, $N{=}1$ Euclidean chiral superfields can be chosen 
as real with respect to ${}*$ (like the general superfields).

Though our main goal is to introduce consistent {\it nilpotent\/} 
deformations of $N{=}(1,1)$ {\it harmonic\/} superspace, it is convenient 
to start the analysis in the standard $N{=}(1,1)$ superspace in the 
{\it chiral\/} parmetrization
\be
z_\sL\ \equiv\ ( x^m_\sL, \tka, \btka )\,.
\ee
These coordinates transform under $N{=}(1,1)$ supersymmetry as
\be
\delta_\epsilon  x^m_\sL\=2\ii(\sigma^m)_\ada\tka\bar\epsilon^{\da k}\ ,
\qq\delta_\epsilon \tka\=\epsilon^\alpha_k\ ,\qq
\delta_\epsilon \btka\=\bar\epsilon^{\da k}\ ,
\lb{standSUSY}
\ee
where $\epsilon^\alpha_k$ and $\bar\epsilon^{\da k}$ are the 
transformation parameters. The `central' bosonic coordinate $x^m$ is 
related to the `left' coordinate by
\be
x^m_\sL\=x^m+\ii(\sigma^m)_\ada\tka\btka\ .
\lb{xLx}
\ee
As automorphisms we have the Euclidean space spinor group $Spin(4)$ and 
the R-symmetry group SU(2)$\times\,$O(1,1)  acting simultaneously on
left and right spinors.

Let us dwell in some detail on the (pseudo)conjugation properties of the 
$N=(1,1)$ superspace. We can assume the Grassmann coordinates to be real 
with respect to the standard conjugation
 \be
\widetilde{\tka}\=\varepsilon^{kj}\varepsilon_\ab\theta_j^\beta\ ,\qq
\widetilde{\btka}\=-\varepsilon_{kj}\varepsilon_{\da\db}
\bar\theta^{\db j}\ ,\qq
\widetilde{x^m_\sL}\={x^m_\sL}\ ,\qq
\widetilde{AB}\=\widetilde{B}\widetilde{A}\ .
\lb{ELconj}
\ee
This conjugation squares to identity on any object, and with respect to 
it the  $N{=}(1,1)$ superspace has the real dimension $(4|8)$.
The component spinor fields have the analogous conjugation properties.
It is evidently compatible with both $Spin(4)$ and R-symmetries, 
preserving any irreducible representation of these groups. However, the 
$N{=}(\ha,\ha)$ superspace cannot be treated as a real subspace of the 
$N{=}(1,1)$ superspace if one considers only this standard conjugation.

Surprisingly, in the same  Euclidean $N{=}(1,1)$ superspace one can 
define an analog of the pseudoconjugation \p{Oo}
\be
(\tka)^*\=\varepsilon_\ab\theta^\beta_k\ ,\qq
(\btka)^*\=\varepsilon_{\da\db}\bar\theta^{\db k}\ ,\qq
(x^m_\sL)^*\=x^m_\sL\ ,\qq
(AB)^*\=B^* A^*
\lb{sr2conj}
\ee
with respect to which the $N{=}(\ha,\ha)$ superspace forms a real 
subspace. The existence of this pseudoconjugation does not imply any 
further restriction on the $N{=}(1,1)$ superspace. It preserves 
representations of $N=(1,1)$ supersymmetry and, like \p{Oo}, is 
compatible with the action of the group $Spin(4)$.
It is also compatible with the R-symmetry group O(1,1). As for the 
R-symmetry group SU(2), it preserves only some U(1) subgroup of the 
latter. In other words, the standard conjugation \p{ELconj} and the
pseudoconjugation \p{sr2conj} act differently on the objects transforming 
by non-trivial representations of this SU(2), e.g. on Grassmann 
coordinates. The map ${}^*$ squares to $-1$ on these coordinates and 
the associated spinor fields, and to $+1$ on any bosonic monomial or 
field. Only on the singlets of the R-symmetry SU(2), e.g. scalar 
$N=(1,1)$ superfields and R-invariant differential operators, both maps 
act in the same way as the standard complex conjugation.
In particular, the invariant actions are real with respect to both
${}^*$ and ${}^\sim$, despite the fact that the component fields may
have different properties under these (pseudo)conjugations. Clearly, it 
is the pseudoconjugation ${}^*$ which is respected by the reduction 
$N{=}(1,1) \;\rightarrow\; N{=}(\ha,\ha)$. Such a reduction preserves 
the pseudoreality but explicitly breaks the SU(2) R-symmetry.

In chiral coordinates, a {\it chiral nilpotent deformation\/} for 
products of superfields is determined by the following operator,
\bea
P &=&-\sha\olp{}^k_\alpha\, C^\ab_{kj} \orp{}^j_\beta\=
-\sha\overleftarrow{Q}{}^k_\alpha\,C^\ab_{kj}\overrightarrow{Q}{}^j_\beta
\qq\textrm{such that} \nn\\[8pt]
A P B &=& -\sha(A\olp{}^k_\alpha) C^\ab_{kj} (\orp{}^j_\beta B)
\=-\sha(-1)^{p(A)}(\partial^k_\alpha A)C^\ab_{kj}(\partial^j_\beta B)
\nn\\[6pt]
&=& -(-1)^{p(A)p(B)}B P A\ .
\lb{Poper}
\eea
Here, $C^\ab_{kj}=C^\ba_{jk}$ are some constants,
$\;p(A)$ is the supersymmetry $Z_2$-grading,
while $Q^k_\alpha=\partial^k_\alpha$ are the generators of left 
supersymmetry and the derivatives act as
\be
\partial^k_\alpha\tib=\delta_i^k\delta^\beta_\alpha
\qq\textrm{and}\qq
\bar\partial_{\da i}\btkb=\delta^k_i\delta^\db_\da\ .
\ee
By definition, the operator $P$ is nilpotent, $P^5=0$. It preserves both 
chirality and anti-chirality and does not touch the SU$(2)_R$.
It induces a graded Poisson bracket on superfields \cite{FL,KPT}.
We also demand $P$ to be real, i.e.~invariant under some antilinear map
in the algebra of superfields.
The two possible (pseudo)conjugations introduced above then lead to
different conditions
\bea
&& \p{sr2conj} \qq\Longrightarrow\qq (C^\ab_{kj})^*\=C_{\ab kj} \\
&& \p{ELconj} \qq\Longrightarrow\qq \widetilde{C^\ab_{kj}}\=C_\ab^{kj}\ .
\eea
Since $(APB)^*=B^*PA^*$ and $\widetilde{A P B}=\widetilde{B}P
\widetilde{A}$,
our star-product satisfies the following natural  rules:
\be
(A \star B)^* \= B^* \star A^*\ ,\qq
\widetilde{(A \star B)} \= \widetilde B \star \widetilde A\ .
\lb{Pconj}
\ee
Under SU$(2)_L\times\,$SU(2), the constant deformation matrix~C 
decomposes into a (3,3) and a (1,1) part (see also~\cite{KPT,FLM}),
\be
C^\ab_{kj}\=C^{(\ab)}_{(kj)}+\varepsilon^\ab\varepsilon_{kj}I\ .
\lb{const}
\ee
It is worth pointing out that the (1,1) part preserves the full
SO$(4)\times\,$SU(2) symmetry.

Note that the manifestly $N{=}2$ supersymmetric bi-differential operators
of~\cite{FL,KPT} involve flat spinor derivatives $D^k_\alpha$ instead of
partial derivatives. Thus they violate chirality. We basically follow
the line of~\cite{Se} and investigate deformations which preserve 
irreducible representations based on chirality and/or Grassmann harmonic 
analyticity (see Section~3), but may explicitly break some fraction of 
supersymmetry.

Given the operator \p{Poper}, the Moyal product of two superfields reads
\bea
A \star B &=& A\,\e^P B \=
A\,B+ A\,P\,B + \sha A\,P^2 B + \sfrac16 A\,P^3 B + \sfrac{1}{24} A\,P^4 
B \lb{moyal}
\eea
where the identity $P^5=0$ was used.
It is easy to see that the chiral-superspace integral
of the Moyal product of two superfields is not deformed,
\bea
\int\!\diff^4x\, \diff^4\theta\; A \star B \=
\int\!\diff^4x\, \diff^4\theta\; A\, B\ ,
\eea
while integrals of star products of three or more superfields are 
deformed.

In our treatment only free actions preserve all supersymmetries
while interactions get deformed and are not invariant under all standard
supersymmetry transformations. To exhibit the residual symmetries of a 
deformed interacting theory, we formulate the invariance condition
\be
[K, P]\=0
\ee
for the corresponding generators~$K$ in the standard $N{=}(1,1)$ 
superspace. Clearly, this condition is generically not met by 
differential operators depending on~$\tka$ and the symmetries 
generated by these are explicitly broken in the deformed superspace 
integrals. Out of all supersymmetry and automorphism generators, only 
$Q_\alpha^k$ and $\bar L^\da_\db$ do commute with~$P$ of~\p{Poper}.
Hence, for a generic choice of the constant matrix $(C^{\ab}_{ik})$,
the breaking pattern is $N{=}(1,1)\to N{=}(1,0)$ for supersymmetry and
SO$(4)\times\,$O$(1,1)\times\,$SU$(2)\,\to\,$SU$(2)_R$
for Euclidean and R-symmetries.

An exception occurs for the singlet part in \p{const}, i.e. for
\be
C^{(\ab)}_{(kj)}\=0 \qq\Longrightarrow\qq
C^\ab_{kj}\=\varepsilon^\ab\varepsilon_{kj} I
\qq\Longleftrightarrow\qq
P_s\=-\sha\,\overleftarrow{Q^k_\alpha}\,I\,\overrightarrow{Q^\alpha_k}
\ , \lb{Cinv}
\ee
which is fully SO$(4)\times\,$SU(2) invariant and non-degenerate but also
fully breaks the right half of supersymmetry.

It is worth noting that it is possible to break less than one half of 
the supersymmetry if we choose the ${}^*$ conjugation \p{sr2conj}.
This choice is compatible with the decomposition of $N{=}(1,1)$ into two
$N{=}(\ha,\ha)$ superalgebras, each given by a fixed value for the 
SU(2)~index. Therefore, it allows one to pick a degenerate deformation, 
e.g.
\be
P_{deg}(Q^2)
\=-\sha C^{12}_{22} (\overleftarrow{Q}{}^2_1\,\overrightarrow{Q}{}^2_2
+ \overleftarrow{Q}{}^2_2\,\overrightarrow{Q}{}^2_1)\ ,\lb{1break}
\ee
which does not involve $Q^1_\alpha$.
In this case, only $\bar Q_{\da 2}$ are broken but not the supercharges
$\bar Q_{\da 1}$. Hence, the deformation $P_{deg}$
preserves $N{=}(1,\ha)$ supersymmetry.

\vspace{0.3cm}
\noindent{\bf 3. Deformations of N=(1,1) harmonic superspace}.
The basic concepts of the $N{=}2, D{=}4$ harmonic superspace \cite{GIK1}
are collected in the book \cite{GIOS}. The spinor $SU(2)/U(1)$ harmonics
$u^\pm_i$ can be used to construct analytic coordinates
$(x_\A^m,\theta^{\pm\alpha},\bar\theta^{\pm\da}, u^\pm_i)$ in the 
Euclidean version of $N{=}2$ harmonic superspace, that is $N{=}(1,1)$ 
harmonic superspace:
\bea
&& x^m_\A\=
x^m_\sL-2\ii(\sigma^m)_\ada\theta^{\alpha k}\btja u^{-}_ku^+_j\ ,\qq
\theta^{\alpha\pm}\= \theta^{\alpha k}u^\pm_k\ ,\ \q
\bar\theta^{\pm\da} \= \bar\theta^{\da k}u^\pm_k
\lb{ancoor}
\eea
where $\epsilon^{\pm\alpha}{=}\epsilon^{\alpha k}u^\pm_k~,~
\bar\epsilon^{\pm\da}{=}\bar\epsilon^{\da k}u^\pm_k~$, and
$(x^m_\sL,\tka,\btka)$ are chiral coordinates of $N{=}(1,1)$ superspace. 
We extend the (pseudo)conjugations  \p{ELconj} and \p{sr2conj} to the 
harmonics by
\be
\widetilde{u^\pm_k}\=u^{\pm k} \qq\textrm{and}\qq  (u^\pm_k)^*\=u^\pm_k 
\label{Ooo}
\ee
so that the analytic coordinates are conjugated identically for both 
choices,
\bea
 &&\widetilde{x_\A^m}\=x_\A^m\ ,\qq
\widetilde{\theta^{\pm\alpha}}\=\varepsilon_{\ab}\theta^{\pm\beta}\ ,\qq
\widetilde{\bar\theta^{\pm\da}}\=\varepsilon_{\da\db}\bar\theta^{\pm\da}\
 , \\[4pt]
 &&(x_\A^m)^*\=x_\A^m\ ,\qq
(\theta^{\pm \alpha})^*\=\varepsilon_{\ab}\theta^{\pm\beta}\ ,\qq
(\bar\theta^{\pm\da})^*\=\varepsilon_{\da\db}\bar\theta^{\pm\da}\ ;
\eea
in particular, both square to $-1$ on spinor coordinates. This means 
that both maps become pseudoconjugations when applied to the extended set
of coordinates. These two pseudoconjugations act identically on 
invariants and harmonic superfields, e.g.~$(A^kB_k)^*=\widetilde{(A^k
B_k)}$ or $(q^+)^*=\widetilde{q^+}$, but differ on harmonics or R-spinor 
component fields, e.g.~$(A_k)^*\neq\widetilde{A_k}$. An important 
invariant pseudoreal subspace is the analytic Euclidean harmonic 
superspace, parametrized by the coordinates
\be
\zeta \ \equiv\ (x_\A^m~,~\theta^{+\alpha}~,~\bar\theta^{+\da}, u^{\pm i})\
.
\ee

The explicit form of supersymmetry-preserving spinor and harmonic 
derivatives in these coordinates can be found in \cite{GIOS}.
The partial derivatives in different bases are related as
\bea
&&\partial_m^\sL\=\partial_m^\A\ , \qq
D^{++}_\sL \= \partial^\pp \= D^{++}_\A\ , \nn\\[6pt]
&&\partial^k_\alpha\=-u^{+k}\partial^-_\alpha-u^{-k}\partial^+_\alpha
+2\ii u^{-k}\btpa(\sigma^m)_\ada\partial^\A_m\ , \nn\\[6pt]
&&\bar\partial_{\da k}\=u^+_k\bar\partial^-_\da+u^-_k\bar\partial^+_\da
+2\ii u_k^+\tma(\sigma^m)_\ada\partial^\A_m
\= u^-_k\bDpa-u^+_k\bDma
\lb{pthk}
\eea
where $\partial^\pm_\alpha\equiv\partial/\partial\theta^{\mp\alpha}$,
$\bar\partial^\pm_\da\equiv\partial/\partial\bar\theta^{\mp\da}$,
$\partial^\pp = u^{+i}\partial/\partial u^{-i}$.
A Grassmann analytic superfield is defined by
\be
\Dpa \Phi(\zeta, \theta^-, \bar\theta^-, u) \=
\bDpa \Phi(\zeta, \theta^-, \bar\theta^-, u) \= 0
\ee
and so can be treated as an unconstrained function in the analytic 
superspace, $\Phi=\Phi(\zeta,u)$.

It is important to realize that the chirality-preserving operator~$P$
in~\p{Poper} also preserves {\it Grassmann analyticity\/}.
This is seen in the analytic basis using the relations \p{pthk},
\be
\{\partial^k_\alpha, \Dpb \} \= \{\partial^k_\alpha, \bDpb \} \=0
\qq\Longrightarrow\qq [P,\Dpb] \= [P,\bDpb] \= 0 \ .
\ee
For the singlet deformation \p{Cinv} we have the following
deformation operator
\bea
&&
P_s\=-\ii (\sigma^m)^\adb \btp_\db\,I\,(
\overleftarrow{\partial^\A_m}\, \overrightarrow{\partial^-_\alpha} -
\overleftarrow{\partial^-_\alpha}\, \overrightarrow{\partial^\A_m} )
\eea
which satisfies $P_s^3=0$.

If we do not care about chirality we may add to $P$ any one of
the two supersymmetry-preserving operators which in analytic
coordinates read
\be
L \= \sha \bigl(\olD{}^{+\alpha} J \orD{}^{-}_\alpha + 
\olD{}^-_\alpha  J \orD{}^{+\alpha}\bigr)\ , \qq
R \= \sha \bigl(\olbD{}^{+\da} \bar{J} \orbD{}^-_\da + 
\olbD{}^-_\da \bar{J} \orbD{}^{+\da}\bigr)\ .
\lb{Lanal}
\ee
It is straigtforward to see that these operators indeed do not preserve
one of the chiralities (in contrast to the operators $P$ which preserves 
both ones). They strongly preserve harmonic analyticity, not deforming 
at all products of analytic superfields $\Phi(\zeta, u)$ and 
$\Lambda(\zeta, u)$:
\be
\Phi\,\e^L \Lambda \= \Phi\,\Lambda\ , \qq
\Phi\,\e^R \Lambda \= \Phi\,\Lambda\ .
\ee
Note that the superfield geometry of gauge theories in the deformed
harmonic superspace with the deformation operator $L$ was studied in 
\cite{FS}.

\vspace{0.3cm}
\noindent{\bf 4. Interactions in deformed harmonic superspace}.
Harmonic superspace with noncommutative bosonic coordinates $x^m_\A$ has 
been discussed in~\cite{BS}. This deformation yields nonlocal theories
but preserves the whole $N{=}2$ supersymmetry. In contrast, we expect 
that the deformations of $N{=}(1,1)$ superspace defined in the previous 
section will produce much weaker nonlocalities due to their nilpotency. 
Leaving quantum considerations for future study, we present here the 
chirally deformed versions of the off-shell actions for some basic 
theories in harmonic superspace.

We shall limit our attention to the deformation operator $P$ which 
affects analytic superfields and preserves both analyticity and 
chiralities, while breaking at least one quarter of $N{=}(1,1)$ 
supersymmetry. The free $q^+$ and $\omega$ hypermultiplet actions of 
ordinary harmonic theory \cite{GIOS} are not deformed in 
non\-(anti)\-commutative superspace:
\be
S_0(q^+)\=\int\!\diff u\,\diff\zeta^{-4}\ \tilde{q}{\,}^+ \Dp q^+\ ,\qq
S_0(\omega)\=\int\!\diff u\,\diff\zeta^{-4}\,(\Dp\omega)^2 \lb{freeAc}\ ,
\ee
where $\diff\zeta=\diff^4x_\A (D^-)^4$.
Non\-(anti)\-commutativity arises in interactions, for instance for 
the self-interactions of the hypermultiplet which contain higher-order 
terms of the type $\sim \tilde{q}{}^+ \star q^+ \star \tilde{q}{}^+
\star q^+ $. Expanding out the star products yields a finite number of 
corrections to the local interaction term $(q^+\tilde{q}{}^+)^2$.

The interaction of the hypermultiplet~$q^+$ with a U(1) analytic gauge
superfield~$\Vp$ can be introduced as in \cite{BS}, by replacing $\Dp$
in~\p{freeAc} with the covariant harmonic noncommutative left-derivative,
\be
\Dp q^+ \qq \Longrightarrow \qq \nabla^\pp q^+\=\Dp q^+ + \Vp \star q^+ .
\lb{defD}
\ee
The gauge transformation of the anti-Hermitian $\Vp$ reads
\be
\delta_\lambda\Vp\=-\Dp \lambda+[\lambda \stc \Vp]
\ee
where $\lambda$ is an anti-Hermitian analytic gauge parameter.
The generalization to U($n$) analytic gauge fields is straightforward.
Note again that from the beginning we retain only those symmetries
which are unbroken by the deformation of choice.

In Wess-Zumino gauge we have
\bea
\Vp_{\W\Z} &=& (\tp)^2\bar\phi\ +\ (\btp)^2\phi\ +
2\tpa\btpa A_\ada \nn\\[4pt]
&+& 4(\tp)^2\btpa u^-_k \bar\lambda^k_\da\ +
4(\btp)^2\tpa u^-_k\lambda^k_\alpha\ +3 (\tp)^2(\btp)^2u^-_ku^-_jD^{kj}\ ,
\lb{WZ}
\eea
with all components being functions of~$x^m_\A$,
and a component expansion of the hypermultiplet~$q^+$ which consists
of infinitely many terms due to the harmonic dependence.
The component expansion of the deformed products is rather complicated
since the number of terms increases significantly.
E.g. for the singlet deformation $P_s$, the star product in~\p{defD}
contains the terms $\Vp P_s\,q^+$ and $\Vp P_s^2\,q^+$.

The action for this noncommutative U(1) gauge superfield can be 
constructed in central coordinates in analogy with the action for 
commutative $N{=}2$ Yang-Mills theory~\cite{Z1}, but it is easier to 
analyze it in chiral coordinates. Following~\cite{Z1}, one constructs 
the deformed connection for the derivative $\Dm$ via
\bea
&&\Dp\Vm-\Dm\Vp+[\Vp \stc \Vm]\=0\ ,\\[6pt]
&&\Vm(z_\sL, u)\=
\sum\limits_{n=1}^\infty(-1)^n\!\int\!\diff u_1\ldots\diff u_n
\frac{\Vp(z_\sL, u_1) \star \Vp(z_\sL, u_2) \ldots \star \Vp(z_\sL, u_n)}
{(u^+u^+_1)(u^+_1u^+_2)\ldots (u^+_nu^+)}\ ,\nn
\eea
where $(u^+_1u^+_2)^{-1}$ is a harmonic distribution (see \cite{GIOS}).
In general, the action for~$\Vp$ contains an infinite number of vertices,
with star commutators substituting the ordinary commutators of~$\Vp$
taken from the standard non-Abelian action.
The chiral and anti-chiral superfield strengths $\cW$ and $\bcW$
in the Euclidean case are independent. They have the form
\be
\cW\=-{1\over4}(\bar D^+)^2\Vm\ , \q \bcW=-{1\over4}(D^+)^2\Vm\ ,\q
\textrm{with}\q\delta_\lambda\,(\cW, \bcW)\=[\lambda \stc (\cW, \bcW)]\ ,
\ee
and satisfy the covariantized chirality and harmonic-independence 
conditions
\be
\bDpa\cW\ = 0\ ,\q
\bDma\cW-[\bDpa\Vm \stc \cW]\=0\ ,\q
\Dp\cW + [\Vp \stc \cW] \=0\ ,
\lb{covharm}
\ee
plus analogous conditions on the anti-chiral $\bcW$,
as well as $\ (D^+)^2\cW = (\bar D^+)^2\bcW$. For the case of the 
chirality-preserving deformations, one can write down
gauge-invariant actions holomorphic in~$\cW$, such as
\be
S_{\cW}\ \sim\ \int\!\diff^{4}x_\sL\,\diff^4\theta\
(\cW^2+a\cW \star \cW \star\cW) , \lb{Abel1}
\ee
where $a$ is some constant. It is easy to check that
\be
\delta_\lambda\, S_{\cW} \= 0 \qq\textrm{and}\qq
\Dp\, S_{\cW} \= \bar D_{\da k}\,S_{\cW} \= 0\ .
\ee

In the Feynman rules, the only effect of our deformations is a small 
number of higher-derivative contributions to the standard interaction 
vertices. Due to the nilpotency of these deformations, the locality of 
the theory is not jeopardized. It should be straightforward to evaluate 
the ensueing mild corrections to the known quantum properties of 
$N{=}(1,1)$ harmonic superspace.

\vspace{0.3cm}
\noindent{\bf 5. Conclusions}. We have considered nilpotent deformations 
of $N{=}(1,1)$ chiral and Grassmann-analytic harmonic superfields, such 
that only the anticommutator of half of fermionic coordinates is deformed 
in a chiral basis. We focussed on those deformations which preserve both 
chirality and harmonic analyticity, but break $N{=}(1,1)$ supersymmetry: 
either to $N{=}(1,0)$ or to $N{=}(1,\ha)$ (for the degenerate
deformation matrix and ${}^*$ conjugation). The second opportunity exists
contrary to an assertion of \cite{last}. On the background of 
non-deformed Euclidean $N{=}(1,1)$ superspace, one can treat such 
deformations as a soft breaking of the part of supersymmetry and 
automorphism symmetry. Complete supersymmetry can only be saved at the 
expense of chirality, though with preserving harmonic analyticity. We 
gave examples of superfield theories in chiral-nilpotently deformed 
harmonic superspace. In particular, we have shown how to construct the 
SO(4)$\,\times\,$SU(2) invariant nilpotent deformation of superfield 
$N{=}(1,1)$ supersymmetric U(1) gauge theory in chiral coordinates.

It would be interesting to understand a possible stringy origin
of the deformations considered here and in \cite{FS} and to work out the 
component form of the deformed superfield actions. For the $N{=}(1,1)$ 
supersymmetry-preserving nilpotent deformation of $N{=}(1,1)$ SYM theory 
the component action was given in \cite{FS}; an analogous consideration 
for the chirality-preserving SO(4)$\,\times\,$SU(2) invariant deformation 
is now under way \cite{FILSZ}.

\vspace{0.3cm}
\noindent{\bf Acknowledgements}.
This work was partially supported by the DFG priority program SPP 1096 
in string theory, INTAS grant No 00-00254, RFBR-DFG grant No 02-02-04002,
grant DFG No 436 RUS 113/669, RFBR grant No 03-02-17440 and a grant of 
the Heisenberg-Landau program.

\end{document}